\documentclass[twocolumn, prb, showpacs, english, amsmath, amssymb, superscriptaddress, aps,longbibliography]{revtex4-2}

\usepackage[utf8]{inputenc}
\usepackage{graphicx}
\usepackage[usenames,dvipsnames]{xcolor}
\usepackage{amsmath}
\usepackage[resetlabels, labeled]{multibib}
\newcites{S}{References Supplementary Materials}
\definecolor{orange}{rgb}{1,0.5,0}
\definecolor{goodgreen}{rgb}{0.1,0.5,0}
\definecolor{goodred}{rgb}{0.7,0,0}
\usepackage[colorlinks,urlcolor=goodgreen,citecolor=blue,linkcolor=goodred]{hyperref}
\usepackage{lineno}
\usepackage{bm,dsfont}

\usepackage{tikz}
\usepackage{tocvsec2}
\usepackage{color,soul}

\newcommand{\ci}{\mathrm{i}}
\newcommand{\bea}{\begin{eqnarray}}
\newcommand{\eea}{\end{eqnarray}}
\newcommand{\be}{\begin{equation}}
\newcommand{\ee}{\end{equation}}

\usepackage[normalem]{ulem}

\usepackage{float}
\makeatother

\usepackage{babel}

\begin{document}

\title{Magnetic switching of spin-scattering centers in Dresselhaus [110] circuits}

\author{E. J. Rodr\'{\i}guez}
\email{erfernandez@us.es}
\affiliation{Departamento de F\'isica Aplicada II, Universidad de Sevilla, E-41012 Sevilla, Spain}

\author{A. A. Reynoso}
\email{areynoso@us.es}
\affiliation{Departamento de F\'isica Aplicada II, Universidad de Sevilla, E-41012 Sevilla, Spain}
\affiliation{Instituto Balseiro and Centro At\'omico Bariloche, Comisi\'on Nacional de Energ\'ia At\'omica, 8400 Bariloche, Argentina}

\author{J. P. Baltan\'as}
\email{baltanas@us.es}
\affiliation{Departamento de F\'isica Aplicada II, Universidad de Sevilla, E-41012 Sevilla, Spain}

\author{J. Nitta}
\email{nitta@material.tohoku.ac.jp}
\affiliation{Department of Materials Science, Tohoku University, Sendai, 980-8579, Japan}
\affiliation{NTT Basic Research Laboratories, Atsugi, 243-0198, Japan}

\author{D. Frustaglia}
\email{frustaglia@us.es}
\affiliation{Departamento de F\'isica Aplicada II, Universidad de Sevilla, E-41012 Sevilla, Spain}
\begin{abstract}

Spin carriers subject to Dresselhaus [110] (D110) spin-orbit coupling (SOC) gather null spin phases in closed circuits, contrary to usual Rashba and Dresselhaus [001] SOC. We show that D110 spin phases can be activated in square circuits by introducing an in-plane Zeeman field, where localized field inhomogeneities act as effective spin-scattering centers. Our simulations show rich interference patterns in the quantum conductance, which work as maps for a geometric classification of the propagating spin states. We also find that disorder facilitates
low-field implementations.
\end{abstract}
\maketitle

\section{Introduction} 

Spin-orbit coupling (SOC) in two-dimensional electron gases (2DEG) \cite{Winkler03} is a strategic resource for quantum electronics and spin-based technologies \cite{reviewBercioux15,reviewNitta15,Schliemann17,reviewHirohata20}. The cases of Rashba \cite{Rashba60} and Dresselhaus [001] \cite{Dressel55} SOC in zincblende III-V compound semiconductor quantum wells have been discussed extensively in the literature over the last decades. Although they belong to different symmetry classes, both Rashba and Dresselhaus [001] SOC present in-plane effective field textures exploited in, e.g., Aharonov-Casher (AC) spin interferometry \cite{Aharonov1984,Mathur1992,Nitta1999,Frustaglia2004,Molnar2004,Bergsten2006,Konig2006,Grbic2007,Aharony11,Nagasawa2012,Nagasawa2018,Lia21,Shekhter22} and the manipulation of geometric spin phases in electronic transport \cite{Aronov1993,Frustaglia2004,Nagasawa2012,Nagasawa2013,Frustaglia20} by electrical control of the corresponding SOC strengths \cite{Nitta1997,Deet17,Nagasawa2018}. By contrast, Dresselhaus [110] (D110) SOC \cite{Dressel55} 
has received relatively little attention
(with some notable exceptions \cite{Winkler03,Hassenkam97,Ohno99,Couto07,Liu2010,Ganichev2014,Schliemann17,Iizasa2018,You2013}). 
However, D110 offers a unique feature of practical interest: an effective field texture perpendicular to the 2DEG's plane. Such a robust symmetry facilitates the development of a so-called persistent spin helix (PSH) \cite{Iizasa2018,Chen2008,Bernevig2006} and the corresponding suppression of spin relaxation \cite{Ohno99}, spin dephasing \cite{Poshakinskiy2013}, spin Hall effect \cite{Flatte06}, and weak antilocalization \cite{Hassenkam97} for conduction electrons. One additional consequence of utmost importance is the absence of AC interference in D110 systems, as in any PSH. 
This is true not only for III-V compound semiconductor 2DEGs but also for group III and IV monochalcogenide monolayers.
Moreover, D110 has been proposed for realizing topological phases in noncentrosymmetric superconductors that could host the celebrated (and elusive) Majorana modes \cite{You2013}.  

In this article, we discuss D110-based spin interferometry and control in mesoscopic circuits of square shape. Due to its symmetry properties, D110 alone does not manifest any sign of AC interference in electronic transport. We show that the introduction of an additional in-plane Zeeman field activates complex spin dynamics leading to the development of spin phases and interference effects that modulate the quantum conductance. In polygonal circuits, the interplay between D110 and Zeeman fields creates effective-field discontinuities localized at the vertices that produce spin scattering. This results in field-dependent phase differences between counter-propagating spin carriers that produce a rich interference pattern. The effect is remarkable, as it enables the magnetic switching of spin-scattering centers in D110 circuits (difficult to achieve with other SOC classes) and its characterization by means of transport experiments. Furthermore, we show that this result is robust against disorder and realizable at low field strengths thanks to the Al'tshuler-Aronov-Spivak (AAS) effect \cite{Altshuler1981}. 

Here, we focus on the modelling of square D110 circuits built on 2DEG lodged in III-V semiconductor heterostructures. Other geometries may work as well; still, we choose square circuits based on its simple experimental realization, modelling, numerical simulation, and interpretation. Previous works have shown that rich conductance patterns arise in Rashba squares \cite{Koga2006,Qu2011,Wang2019}. Moreover one-dimensional (1D) models have been applied with success to Rashba squares \cite{Bercioux2005,Veen06,Hijano2021,Rodriguez2021,Hijano22}, resulting in excellent agreement with the experimental observations. One-dimensional models are especially suited to large loop arrays \cite{Bergsten2006,Koga2006,Nagasawa2012,Nagasawa2013,Nagasawa2018,Wang2019}, where higher slow-propagating modes are more likely to decohere than lower fast-propagating ones. The surviving interference observed in experiments with loop arrays is well described by the physics of one single (quasi 1D) orbital mode propagating along a single mesoscopic loop \cite{Nagasawa2012,Nagasawa2013,Nagasawa2018,Wang2019,Frustaglia20}.

\begin{figure}[!tt]
\centering
    \includegraphics[width=0.45\textwidth]{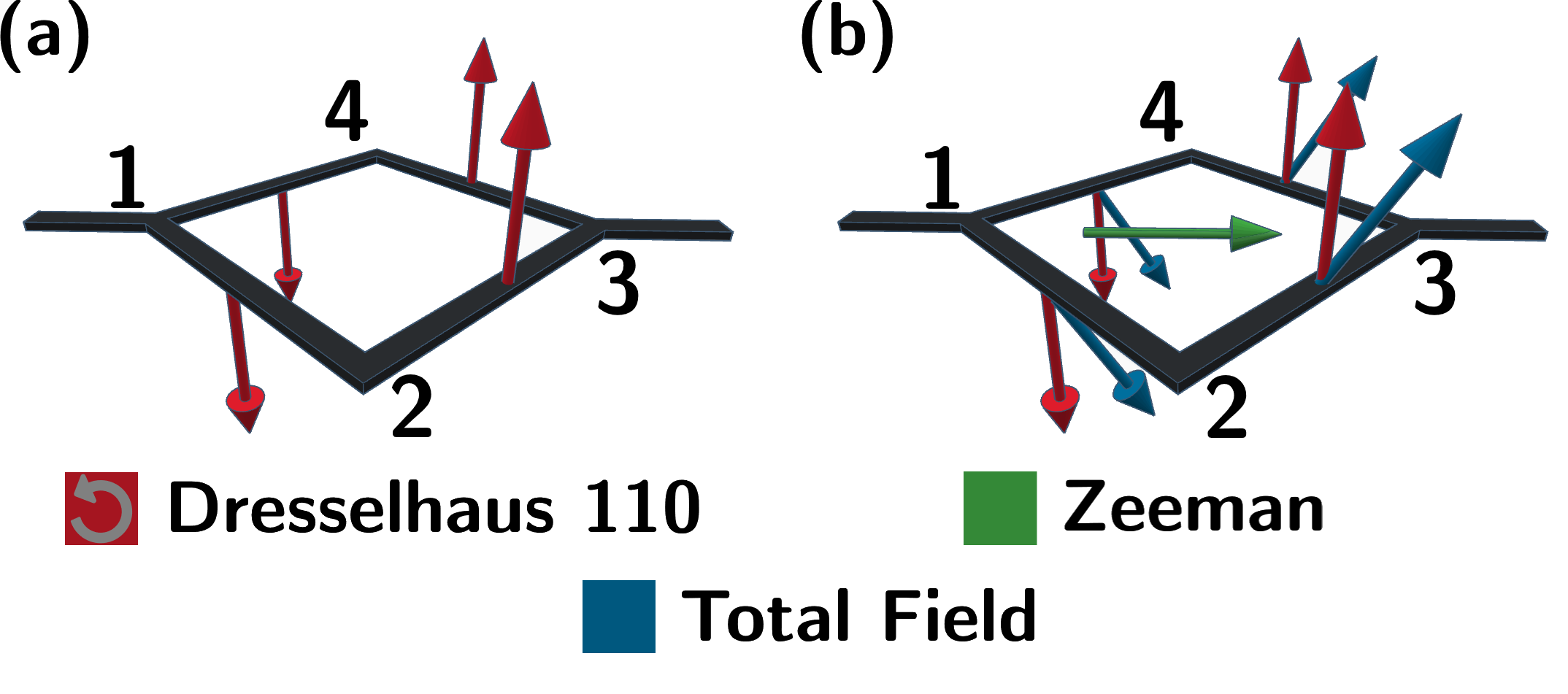}
    \caption{\label{fields} 
    (a) Square circuit subject to D110 SOC. The arrows represent the SOC field experienced by CCW propagating spin carriers. The field lies in the vertical direction, pointing up or down depending on the carriers' $p_y$ momentum at each circuit's section. For CW propagating carriers, the SOC field inverts its sign due to time reversal symmetry. (b) The introduction of a Zeeman field activates discontinuities in the effective-field direction localized at vertices $2$ and $4$.}
\end{figure}

\section{Model} 

We consider a 1D square circuit contained in the $xy$-plane consisting of conducting segments of length $L$ subject to D110 SOC and Zeeman coupling, see Fig.~\ref{fields}. Each segment connects vertices $u$ and $v$ ($u,v = 1,\dots,4$) and is oriented along the direction $\hat{\boldsymbol{\gamma}}=(\cos\gamma, \sin\gamma, 0)$, from $u$ towards $v$. The spin-carrier dynamics along each segment is given by the Hamiltonian \cite{Winkler03,Liu2010,Ganichev2014,Schliemann17,Iizasa2018}: 
\begin{equation}
\label{Hamiltonian}
\hat{\mathcal{H}}_{vu}=\frac{p^{2}}{2m^*}+\frac{\beta}{\hbar}p_y\sigma_z+\mu\boldsymbol{B}\cdot\boldsymbol{\sigma},
\end{equation}
where $p$ is the carrier's linear momentum along $\hat{\boldsymbol{\gamma}}$, $\beta$ is the (renormalized) linear D110 SOC strength, $m^*$ the carrier's effective mass, $\mu$ is the magnetic moment, and $\boldsymbol{\sigma}$ the vector of Pauli matrices. Notice that the D110 SOC contribution |2nd term on the r.h.s. of Eq.~\eqref{Hamiltonian} |appears as a momentum-dependent magnetic field of magnitude $(\beta/\hbar\mu)p_y$ coupled to the spin along the $z$ direction. Importantly, since $p_y=p\sin\gamma$, the orientation of each segment modulates the amplitude and sign of the D110 SOC strength. 
This term establishes a PSH
\cite{Iizasa2018,Chen2008,Bernevig2006}.
There is experimental evidence \cite{Chen14} of a dominating linear-in-momentum contribution to D110 SOC \footnote{Notice that the lesser higher-order contribution proportional to $p^3 \sin3\varphi \ \sigma_z$ \cite{Schliemann17}, with $\varphi$ the polar angle with respect to the $x$ axis, does not compromize the PSH since it acts along the $z$ axis as well as the linear contribution in Eq. (\ref{Hamiltonian}).}. Still, higher-order $p$ contributions can be incorporated by introducing a renormalized $\beta$ in Eq.~\eqref{Hamiltonian} \cite{Schliemann17}. This suggest the possibility of controlling the D110 SOC strength by electrical means, similar to what recently found in Dresselhaus [001] systems \cite{Deet17,Nagasawa2018}.
The in-plane Zeeman field, $\boldsymbol{B}=B \hat{\boldsymbol{b}}_\alpha$, with $\hat{\boldsymbol{b}}_\alpha=(\cos\alpha, \sin\alpha, 0)$, is common to all segments. Hence, the effective field undergone by a spin-carrier travelling either clockwise (CW) or counterclockwise (CCW) around the square loop resembles a stroboscopic Rabi driving, namely, a constant Zeeman field coexisting with a normal D110 component that takes a sequence of discrete values, see Fig.~\ref{fields}(b). Notice that the Zeeman field undermines the PSH established by the D110 SOC field.

The square circuit of Fig.~\ref{fields} is oriented such that, starting from node $1$, the directional angles are $\gamma= \{-\frac{\pi}{4},\frac{\pi}{4},\frac{3\pi}{4},\frac{5\pi}{4}\}$. This choice maximizes the amplitude of the D110 SOC field. The resulting D110 field texture for a spin carrier travelling CCW is shown in Fig.~\ref{fields}(a). It represents a field oscillating along the $z$ direction, fixing a global spin quantization axis. Under this circumstance, the spin phase gathered by a carrier in a round trip is zero. Figure~\ref{fields}(b) shows how the introduction of an external in-plane Zeeman field induces sudden changes in the direction of the effective-field at nodes $2$ and $4$, destroying the global quantization axis. This favors the development of complex spin dynamics and phases due to the misalignment of the local spin quantization axis at different segments. In this way, vertices $2$ and $4$ act as effective spin-scattering centers.
In order to maximize interference effects, in the following we study quantum transport between vertices $1$ and $3$. Any other choice would undermine spin-dependent signatures (in particular, symmetry dictates that flat interference patterns are expected for transport between nodes $2$ and $4$). Still, misalignments up to $30^\circ$ between the leads and the crystallographic axis $x$ are acceptable. See Appendix \hyperref[AppA]{A}.

\begin{figure}[!t]
\centering
    \includegraphics[width=0.45\textwidth]{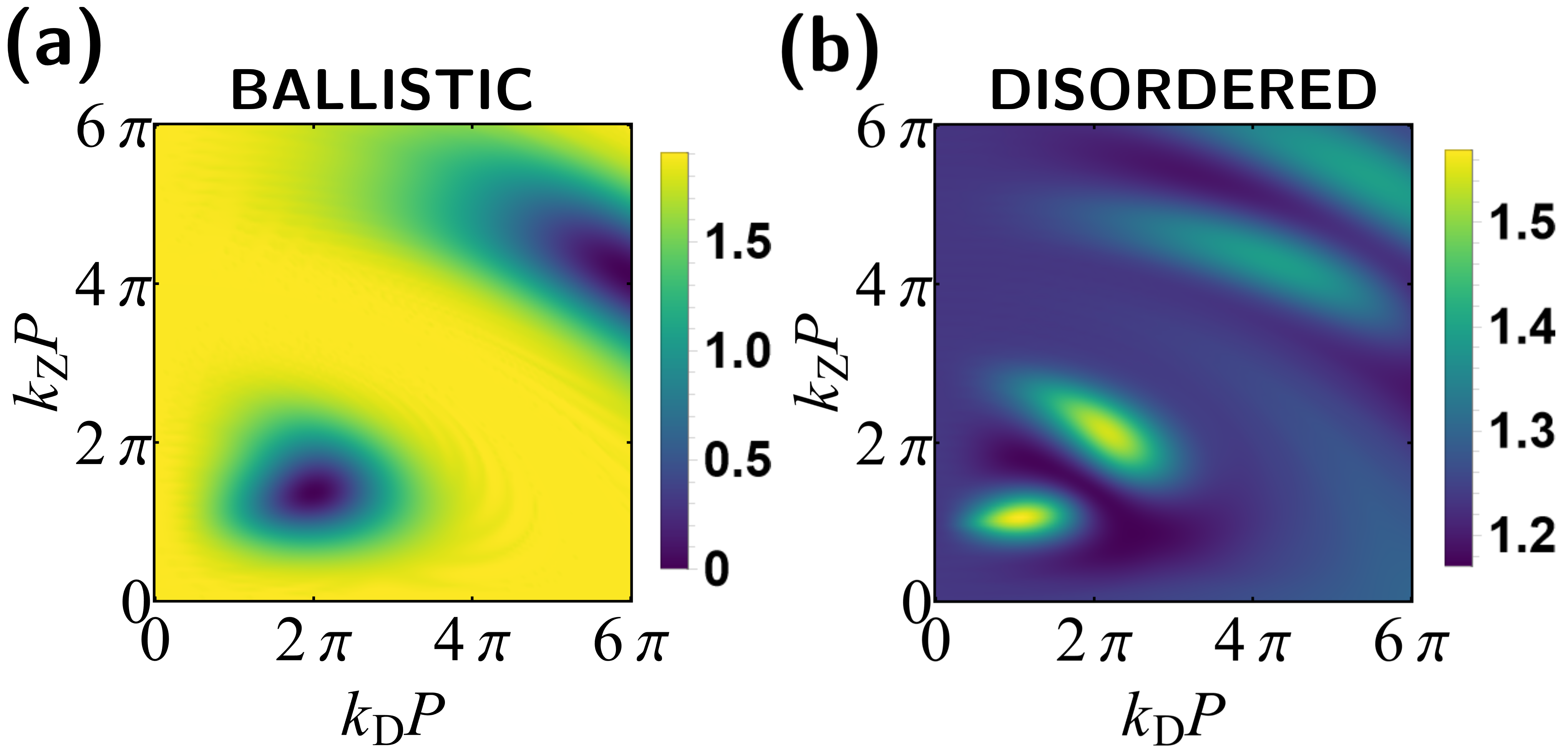}
    \caption{\label{Tight}Tight-binding simulation of the quantum conductance (in units of $e^2/h$) for 1D square loops corresponding to (a) a symmetric circuit in the ballistic regime and (b) a disordered circuit.}
\end{figure}

To study the quantum conductance as a function of D110 and Zeeman fields we implement a tight-binding model of the square loop and use a Green's function formalism. This permits the inclusion of contact leads and accounts for all possible propagation paths contributing to quantum transport between contacts (see Appendix \hyperref[AppB]{B}). Following the Landauer-B\"{u}ttiker formalism \cite{Buttiker1985}, the zero-temperature conductance is given by $G= (e^2/h)T$, with $T$ the quantum transmission between contacts evaluated at the Fermi energy, $E_\mathrm{F}$. 
Notice that a suitable description of the experimental conditions requires an energy average of the computed conductance to discard resonances due to finite-size effects. We focus on the semiclassical limit in which the Fermi wavelength of the electron, $\lambda_\mathrm{F}=2\pi k_\mathrm{F}^{-1}$, with $k_\mathrm{F}=\sqrt{2m^* E_\mathrm{F}}$ the Fermi momentum, is much smaller than the perimeter $P=4L$, i.e., $\lambda_\mathrm{F} \ll P$ \cite{Nagasawa2013,Frustaglia2004}.

\section{Results}

In Figure~\ref{Tight} we show the computed conductance as a function of $k_\mathrm{D}P\equiv (\beta m^*/\hbar^2)P$ and $k_\mathrm{Z}P\equiv(\mu B m^*/\hbar^2 k_\mathrm{F})P$. These parameters are chosen to coincide with the spin phases gathered by a spin carrier propagating along a straight quantum wire of length $P$ oriented along the $y$-axis due to the action of the D110 and Zeeman fields, respectively, with a flight time $P m^*/\hbar k_\mathrm{F}$. Figure~\ref{Tight}(a) shows results for a symmetric square identified with \emph{ballistic} transport through a regular array of square loops. Notice, however, that a typical sample consists of hundreds of loops, so that a more realistic description of the experimental conditions requires a self-averaging over disorder realizations (modelled here by introducing segments of fluctuating lengths taken from a random distribution). We plot the corresponding tight-binding results in Fig.~\ref{Tight}(b). 

Our numerical simulations show that spin interference effects appear as periodic patterns in the quantum conductance over a wide range of field strengths. However, here we focus our attention on relatively low $k_\mathrm{Z}P$ and $k_\mathrm{D}P$ values, according to the limited field strengths accessible in experiments \cite{Nagasawa2013,Wang2019}. In both cases--- ballistic, Fig.~\ref{Tight}(a), and disordered, Fig.~\ref{Tight}(b)--- we observe that the coexistence of D110 and Zeeman fields is essential for the emergence of spin interference effects: D110 or Zeeman fields alone do not imprint any relevant spin phase contributing to interference due to the existence of global spin quantization axes (along $\hat{\boldsymbol{z}}$ and $\hat{\boldsymbol{b}}_\alpha$, respectively), so that the conductance is spin independent. 

In contrast, the interplay between D110 and Zeeman fields produces discontinuities in the 
direction of the local spin-quantization axis at vertices 2 and 4, Fig.~\ref{fields}(b), that turn on spin-scattering processes and spin phases contributing to interference. When both fields are comparable in magnitude, spin interference leads to a periodic series of lobes corresponding to minimum, Fig.~\ref{Tight}(a), or maximum, Fig.~\ref{Tight}(b), conductance. Notice that disordered circuits display split lobes with respect to the ballistic case, with a periodicity of order $2\sqrt{2/3}\pi$ along the diagonal $k_{\rm D}P=k_{\rm Z}P$. The lobe splitting is a manifestation of the AAS effect \cite{Altshuler1981} due to dominant time-reversed interfering paths in disordered loops, also related to weak-(anti)localization in disordered thin layers \cite{bergmann1984weak,Meijer05}. This has dramatic consequences for experimental verification, since it allows for detection of active field discontinuities for $50\%$ lower field magnitudes if one focuses on the first split lobe. 

In-plane magnetic fields of 2.5T lead to $k_{\rm Z}P \approx 2\pi$ in square loops with perimeter $P=2.8 \ \mu$m built on 2DEGs lodged in InGaAs quantum wells (QW) \cite{Wang2019} . Moreover, in single-mode QWs of width $w$ one can approximate $\beta \approx \gamma_{\rm D} (\pi/w)^2$ \cite{Liu2010}, with $\gamma_{\rm D}$ the bulk-inversion asymmetry coefficient running from 11 eV\AA$^3$ in GaAs and InGaAs \cite{Kohda2017} to 490 eV\AA$^3$ in InSb \cite{Kallaher2010}. The QWs can be as narrow as 5-10 nm \cite{Deet17,Liu2010}. Moreover, the carriers' effective mass $m^*$ runs from 0.014 in InSb to 0.023 in InAs, 0.041 in InGaAs, and 0.067 in GaAs (in units of the electron mass $m_0$) \cite{Winkler03}. To be conservative, for square loops with $P=2.8 \ \mu$m built on a $w=10$ nm InGaAs QW we find $k_{\rm D}P\approx \pi$. These values could be tuned on the same sample by building arrays of square loops with different perimeter $P$, to be tested in transport experiments. The electrical tuning of the renormalized $\beta$ (through cubic terms) can also contribute with a tuning range proportional to $\gamma_{\rm D}$ \cite{Nagasawa2018,Deet17}.

We point out that the interference patterns shown in Fig.~\ref{Tight} for D110 squares are essentially different from those reported in Ref. \cite{Wang2019} for Rashba squares. There, the Rashba SOC lead to AC interference patterns subject to additional modulation by an in-plane Zeeman field (with lobes centered on the Rashba axis). Here, instead, no AC interference arises as a consequence of D110 SOC: both D110 and Zeeman fields are necessary for the emergence of interference patterns (with no lobes on the D110 axis).   

Finally, the conductance shows no dependence on the angle $\alpha$, namely, the results of Figs.~\ref{Tight}(a) and \ref{Tight}(b) are valid for any orientation of the in-plane Zeeman field (see Appendix \hyperref[AppC]{C} for a demonstration).

\section{Discussion}

To gain further physical insight on the exact numerical results presented above, we make use of a 1D model that captures the spin dynamics along the loop, disregarding orbital back-scattering at the vertices \cite{Wang2019,Rodriguez2021}. The solutions of the 1D Schr\"odinger equation along a segment based on Eq. \eqref{Hamiltonian} are plane waves such that the spinor wavefunction propagates from vertex $u$ towards vertex $v$ along the direction $\hat{\boldsymbol{\gamma}}$ as 
\begin{equation}
\label{solutionschr}
|\psi(\ell)\rangle=e^{-\mathrm{i} k_{\mathrm{F}} \ell} e^{-i k_{\mathrm{D}} \ell \sin\gamma  \ \sigma_z-i k_{\mathrm{Z}} \ell\hat{\boldsymbol{b}}_\alpha \cdot \boldsymbol{\sigma}}|\psi(0)\rangle,
\end{equation} 
where $\ell$ parametrizes the distance from vertex $u$, with $\ell=L$ at vertex $v$. The prefactor $k_{\mathrm{F}}\ell$ corresponds to the kinetic phase of the carrier associated to the charge dynamics, while the remaining factors represent the spin phase due to spin precession in the presence of D110 and Zeeman fields. Notice that the D110 SOC field inverts its sign for counter-propagating carriers. Indeed, since $\hat{\boldsymbol{\gamma}}$ points from $v$ to $u$, the sign reversal of the D110 contribution is made explicit in Eq.~\eqref{solutionschr} by noting that $\sin(\gamma+\pi)=-\sin\gamma$. 

\begin{figure}[!ht]
\centering
    \includegraphics[width=0.45\textwidth]{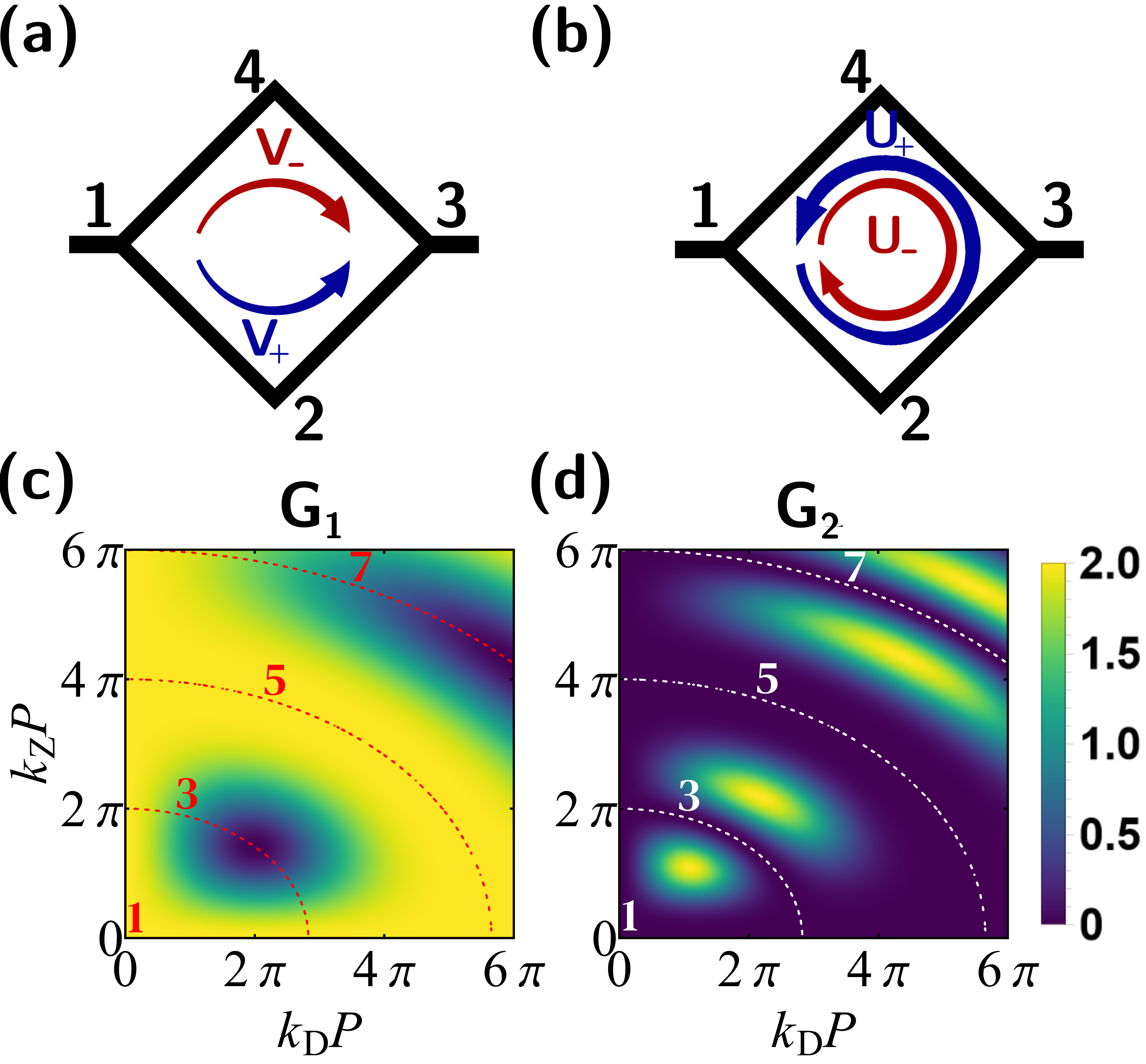}
    \caption{\label{Conductance}(a) Direct paths, $V_{+}$ and $V_{-}$, and (b) time-reversed paths, $U_{+}$ and $U_{-}$, for a square loop. (c) Quantum conductance $G_1$ from direct paths, corresponding to the ballistic case of Fig.~\ref{Tight}(a). (d) Quantum conductance $G_2$ from time-reversed paths, applicable to the disordered case of Fig.~\ref{Tight}(b). The dashed lines indicate field settings such that $U_{\pm}=\mathbb{I}$, also defining winding-number transitions for the spin states. We show sectors with $\omega=1,3,5$ and $7$. The $\omega$ is undefined for vanishing Zeeman or D110 fields.}
\end{figure}

According to Eq.\eqref{solutionschr}, the spin evolution along a full segment is given by the spin rotation operator: \begin{equation}
\label{spinoperator}
R_{v u}=\exp \left[-\mathrm{i} k_{\mathrm{D}} L \sin\gamma \  \sigma_z - i k_{\mathrm{Z}} L\hat{\boldsymbol{b}}_\alpha \cdot \boldsymbol{\sigma}\right].
\end{equation}
By using the labelling shown in Figs.~\ref{Conductance}(a) and \ref{Conductance}(b), we define the spin evolution operators $U_{+}=R_{1 4} R_{43} R_{32} R_{21}$, $U_{-}=R_{12} R_{23} R_{3 4} R_{4 1}$, $V_{+}=R_{32} R_{21}$ and $V_{-}=R_{3 4} R_{4 1}$,
with the subindex $+$ ($-$) indicating CCW (CW) propagation. The unitary $U_{\pm}$ represent full round-trips with origin in vertex 1, whereas $V_{\pm}$ correspond to direct paths from vertex 1 to vertex 3. We resort again to the Landauer-B\"{u}ttiker formalism and obtain the conductance, $G = (e^2 /h)T$, by computing the transmission $T$ for two different dominating interference processes. In the first case, $T$ is computed from the interference between the direct paths $V_{+}$ and $V_{-}$, see Fig.~\ref{Conductance}(a). This contribution is dominant for geometries preserving a twofold reflection symmetry along the axis connecting the contact leads. It requires symmetric and clean samples as, e.g., the ballistic ones discussed in Fig.~\ref{Tight}(a).  In the second case, the transmission is computed as $T=2-R$ after noting that the reflection coefficient $R$ follows from the interference between paths $U_{+}$ and $U_{-}$, see Fig. \ref{Conductance}(b). This contribution dominates in disordered samples, as those discussed in Fig.~\ref{Tight}(b), since $U_{+}$ and $U_{-}$ describe the spin evolution along time-reversed paths. 

Figures \ref{Conductance}(c) and \ref{Conductance}(d) show the conductances $G_1$ and $G_2$ calculated from direct and time-reversed path contributions, respectively, as a function of $k_\mathrm{Z}P$ and $k_\mathrm{D}P$. We find that these lower-order semiclassical models reproduce very well the fully quantum numerical results of Fig.~\ref{Tight} for the ballistic ($G_1$) and disordered ($G_2$) cases. In particular, the agreement between the disordered tight-binding simulations, Fig.~\ref{Tight}(b), and the results from time-reversed path contributions, Fig.~\ref{Conductance}(d), demonstrates that the lobe splitting has its origin in the AAS effect.  

Moreover, an inspection of $U_{\pm}$ unveils relevant geometric characteristics of the spin states. We find that $U_{\pm}$ reduces to the SU(2) identity $\sigma_0$ whenever $\tilde{k}P=2 n \pi$, with $\tilde{k}=\sqrt{k_{\rm Z}^2+k_{\rm D}^2/2}$ and $n$ integer. Notice that the carriers gather no net spin phase in a full round trip for these particular field settings. The condition defines a series of lines that split the lobes in Figs. \ref{Conductance}(c) and \ref{Conductance}(d). Interestingly, we find that such lines also define topological transitions for the textures displayed by the spin modes $|\Psi(\ell)\rangle$ in the Bloch sphere (with $|\Psi(0)\rangle$ an eigen-solution of $U_{\pm}$). This is demonstrated by defining the winding number $\omega$ around the Zeeman-field axis $\hat{\boldsymbol{b}}_\alpha$ as
\begin{equation}
\label{wn}
\omega=\frac{1}{2 \pi} \int_{0}^{P} d \ell\left(\widehat{\boldsymbol{n}} \times \frac{d \hat{\boldsymbol{n}}}{d \ell}\right) \cdot \hat{\boldsymbol{b}}_\alpha,
\end{equation}
where $\hat{\boldsymbol{n}}(\ell)$ is the normalized projection of the spin texture $\hat{\mathbf{s}}(\ell)=\langle\Psi(\ell)|\boldsymbol{\sigma}|\Psi(\ell)\rangle$ on the plane orthogonal to $\hat{\boldsymbol{b}}_\alpha$ (see Appendix \hyperref[AppD]{D} for further details). We find that the spin states organize in sectors with definite odd $\omega$, with only one single lobe per sector in Figs. \ref{Conductance}(d). This finding suggests that the interference pattern in the quantum conductance can work as a map for the geometric characterization of the spin states.

\section{Conclusions and Outlook}

We show that the combined action of D110 SOC and Zeeman fields is a resourceful tool for the manipulation of spin carriers in quantum circuits. This is demonstrated in square circuits as a proof of concept, though other geometries may work as well.  When either field dominates over the other, transport is spin independent (differing from previous results with Rashba squares displaying AC interference \cite{Wang2019}). By contrast, the application of D110 and Zeeman fields of similar magnitude turns on effective spin-scattering centers that trigger complex spin dynamics and the development of spin phases modulating quantum transport. The resulting interference pattern in the conductance works as a guide through the geometry of the propagating spin states. Remarkably, the presence of disorder facilitates the experimental realization in loop arrays at relatively low field strengths due to the AAS effect. This also provides a way to determine the magnitude of the D110 SOC strength in a 2DEG.

Our results provide crucial information about the action of D110 SOC fields on spin carriers and show how spin dynamics can be activated in a controlled fashion, demonstrating a potential for applications in spintronics and spin-based quantum technologies.

We further notice some alternative implementations. Similar effects could be found in usual Rashba and Dresselhaus [001] square circuits (with in-plane effective SOC fields) provided that a PSH is established by setting equal SOC strengths \cite{Schliemann17}. This, however, has the disadvantage of requiring a fine tuning of the SOC fields (contrary to D110 circuits where the PSH is built-in). Moreover, group-III metal-monochalcogenide monolayers such as GaSe and GaS display threefold symmetric, cubic-in-momentum D110 \cite{Li15} which could be optimally exploited by using, e.g., triangular or hexagonal circuits. Recently, it has been suggested \cite{AI2019} that group-IV monochalcogenide $MX$ monolayers ($M=$Sn or Ge and $X=$S, Se, or Te) develop D110 PSHs originated by in-plane ferroelectricity that can be controlled electrically, appearing as possible platforms. An interesting alternative would be electron spin resonance (ESR) experiments in D110 zigzag ballistic channels instead of D110 square loops \cite{Sanada2013}.

\section{Acknowledgements} 

We acknowledge support from the Spanish MICINN-AEI through Project No. PID2021-127250NB-I00 and from the Andalusian Government through PAIDI 2020 Project No. P20-00548 and FEDER Project No. US-1380932. We thank R. Winkler for suggesting us the study of D110 circuits.
\\

\subsection*{Appendix A: Spin Dynamics In Other Configurations}
\label{AppA}
\begin{figure*}[!ht]
\centering
    \includegraphics[width=0.85\textwidth]{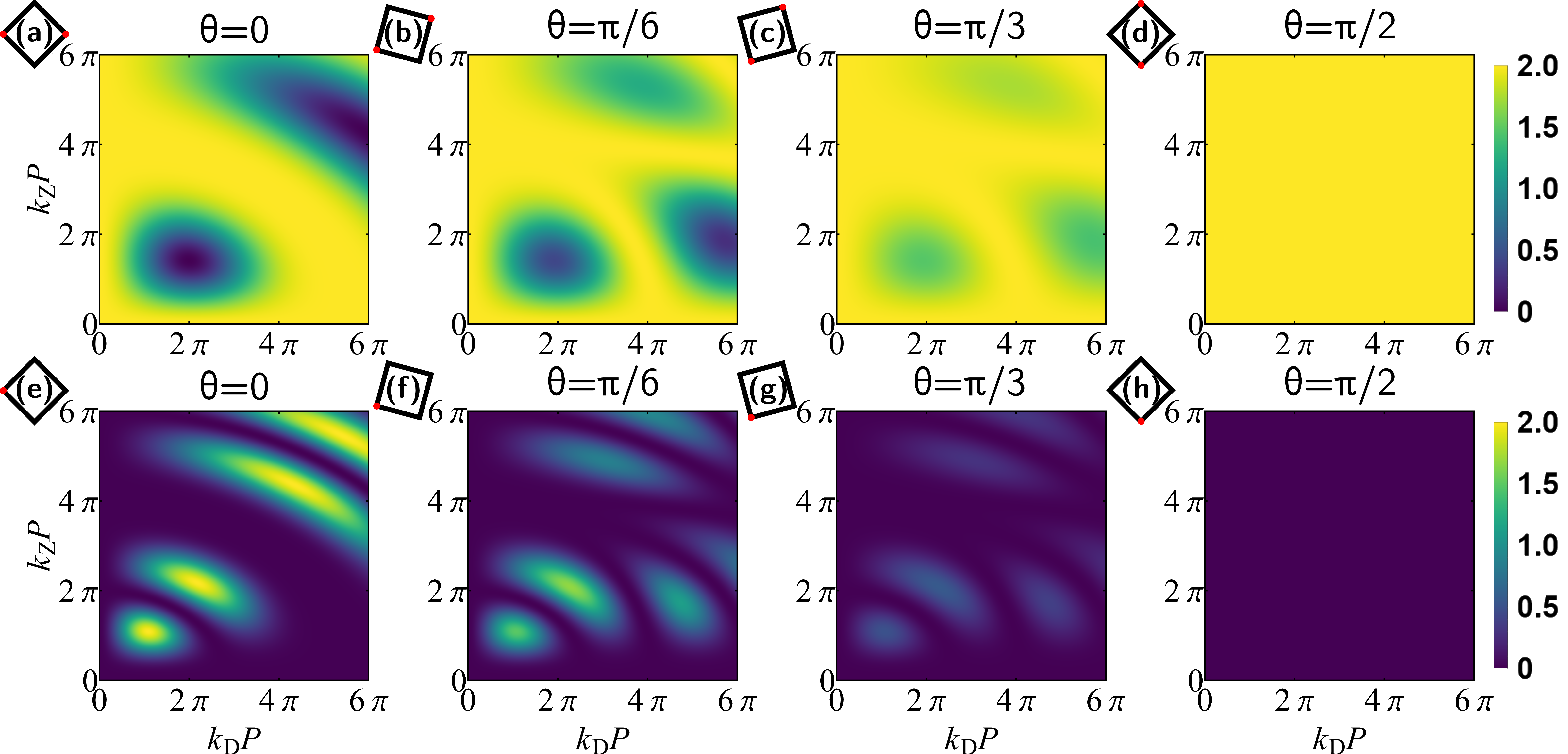}
    \caption{\label{Rotation} Quantum conductance $G_1$ [$G_2$] for sample orientation angle: (a) [(e)] $\theta=0$, i.e., the main-text configuration, (b) [(f)] $\theta=\pi/6$, (c) [(g)] $\theta=\pi/3$, and (d) [(h)] $\theta=\pi/2$. Insets: Our semiclassical model considers spin carriers propagating along paths starting in one red spot and ending in the opposite one for $G_1$ or, for $G_2$, starting and ending at the red spot.}
\end{figure*}

\renewcommand{\theequation}{A.\arabic{equation}}
\setcounter{equation}{0}
Given the orbital anisotropy of the D110 spin-orbit term, the orientation of the sample with respect to the crystallographic axis affects the results. Here we focus on the dependence with the angle of orientation, $\theta$, of the conductances $G_1$ and $G_2$ obtained using the spin-rotation model presented in the main text. In Fig.~\ref{Rotation}, panels (a-d) and (e-h) show results for $G_1$ and $G_2$, respectively, corresponding to $\theta=\{0,\pi/6,\pi/3,\pi/2\}$ as a function of the Zeeman and D110 strengths. The strongest modulation is obtained for $\theta=0$ with contacts in vertices 1 and 3 (sharing the same $y$ coordinate), Fig.~\ref{Rotation} (a) and (e), which corresponds to the configuration studied in the main text. As $\theta$ grows,  interference contrast decreases. For $\theta=\pi/2$, Fig.~\ref{Rotation} (d) and (h), the interference disappears. 

The sharp difference between the $\theta=0$ and $\theta=\pi/2$ cases is a consequence of the D110 strength being proportional to $p_y$. Following Eq.~\eqref{spinoperator} of the main text the spin evolution along a length $L$ and orientation $\gamma$ segment is given by the unitary $R_{v u}=\exp \left[-\mathrm{i} (k_{\mathrm{D}} L \sin\gamma \hat{\mathbf{z}} + k_{\mathrm{Z}} L\hat{\mathbf{x}}) \cdot \boldsymbol{\sigma}\right]$ with the Zeeman field along the $x$ axis. Having nonzero weight in both factors is essential for the transformations on different segments being non commutative, leading to phase differences for different travelled paths. For the conductance $G_1$ the interference is between the two direct paths from vertex $1$ to $3$ (see $V_\pm$ in Fig.~\ref{Conductance}(a) of the main text): either via vertex $2$ or via vertex $4$. For sample orientation $\theta=0$ the first path ($V_+$) is defined by the segment orientations $\gamma=\{-\pi/4,\pi/4\}$, whereas the second ($V_-$) by $\gamma=\{\pi/4,-\pi/4\}$. The two paths lead to $\sin\gamma$ being proportional to $\{-1/\sqrt{2},1/\sqrt{2}\}$ or $\{1/\sqrt{2},-1/\sqrt{2}\}$, and the phase difference controlling the interference depends on the particular values of $k_{\mathrm{D}}$ and $k_{\mathrm{Z}}$. This particular angle difference maximizes such difference leading to the strongest modulation pattern. On the other hand, for sample orientation $\theta=\pi/2$, as the contacted vertex $1$ and $3$ become vertically aligned, the first path has $\gamma=\{\pi/4,3\pi/4\}$, the second path  $\gamma=\{3\pi/4,\pi/4\}$, both leading to $\sin\gamma$ being proportional to $\{1/\sqrt{2},1/\sqrt{2}\}$. This means that both paths produce an identical spin transformation, leading to constructive interference for the quantum transmission from vertex $1$ to $3$, resulting in maximum conductance for any value of $k_{\mathrm{D}}$ and $k_{\mathrm{Z}}$. An analogous analysis can be extended to explain that for $\theta=\pi/2$ the conductance $G_2$ is minimized independently on $k_{\mathrm{D}}$ and $k_{\mathrm{Z}}$: resulting from an identical sequence of $\sin\gamma$ values for both interfering paths generating constructive interference of the return transmission to vertex $1$ (see time-reversal paths $U_\pm$ in Fig.~\ref{Conductance}(b) of the main text). On the other hand, for $\theta=0$, the two paths relevant for $G_2$ have different sequences of $\sin\gamma$, namely,  $\{-1/\sqrt{2},1/\sqrt{2},1/\sqrt{2},-1/\sqrt{2}\}$ and $\{1/\sqrt{2},-1/\sqrt{2},-1/\sqrt{2},1/\sqrt{2}\}$; as shown in Fig.~\ref{Rotation}(d), this is the orientation generating the strongest interference-pattern modulation, induced by the coexistence of $k_{\mathrm{D}}$ and $k_{\mathrm{Z}}$.

\subsection*{Appendix B: Tight-Binding Approach}
\setcounter{equation}{0}
\label{AppB}
\renewcommand{\theequation}{B.\arabic{equation}}
We focus on one-dimensional (1D) circuits of polygonal shape. Each segment of a $N$-sided regular polygon is discretized in $N_s+1$ sites labelled $j=\{0,1,\dots,N_s\}$ and separated by the lattice distance $a_0=P/(N N_s)$. Following the main text, the unit vector $\hat{\mathbf{\gamma}}$ defines the angular orientation for a segment from vertex $u$ to vertex $v$, thus the coordinate for site $j$ is  $\mathbf{r}_j=\mathbf{r}_u+j a_0\hat{\mathbf{\gamma}}$, with $\mathbf{r}_u$ the coordinate of vertex $u$. By applying the customary finite difference method to the segment Hamiltonian of Eq.~(1) one gets the following 1D tight-binding Hamiltonian:
\be
\label{Ht-b}
\begin{gathered}
\hat{\mathcal{H}}_{vu}=
\sum_{j=0}^{N_s}\sum_{\sigma\sigma'}\left(2 t_\mathrm{h}\sigma_0+\mu\boldsymbol{B}\cdot\boldsymbol{\sigma}\right)_{\sigma\sigma'}
\hat{c}_{j\sigma}^\dagger\hat{c}_{j\sigma'}^{}
\\
+\sum_{j=0}^{N_s-1}\sum_{\sigma\sigma'}\left[ \left( \ci t_\mathrm{D} \cos\gamma\sigma_z-t_\mathrm{h} \sigma_0 \right)_{\sigma\sigma'}\hat{c}_{j+1,\sigma}^\dagger\hat{c}_{j\sigma'}^{}+h.c. \right]\,,
\end{gathered}
\ee
where $\sigma_0$ is the SU(2) identity matrix, $\hat{c}_{j\sigma}^{}$ ($\hat{c}_{j\sigma}^\dag$) refers
to the annihilation (creation) operator for an electron in site $j$ with spin $\sigma=\{ \uparrow,\downarrow \}$ along the $z$ direction, $t_\mathrm{h}=\hbar^2/(2 m^* a_0^2)$ is the hopping energy, and $t_\mathrm{D}=\beta/(2 a_0)$ is the D110 hopping energy. The Hamiltonian for the full polygon, $\hat{\mathcal{H}}_{p}$, simply follows from the sum of the Hamiltonians of the $N$ segments (avoiding double counting local terms at the $N$ vertices). 

The system is connected to a source contact lead and a drain contact lead, labelled $\eta=\mathrm{s},\mathrm{d}$, modeled by semi-infinite tight-binding chains parametrized by the hopping energy $t_\eta$ and the site energy $\epsilon_\eta$. The retarded Green function evaluated at the energy $\varepsilon$ for the edge site of each semi-infinite chain is $\hat{g}^r_\eta(\varepsilon)$, having a local and spin-trivial matrix element, $g^r_\eta(\varepsilon)$, which is easily obtained numerically or analytically from the continuous fraction equation $1/g^r_\eta(\varepsilon)=\varepsilon+\mathrm{i}0^+-\epsilon_\eta-t_\eta^2 g^r_\eta(\varepsilon)$. Each lead is connected to a particular vertex of the system (e.g., vertices 1 or 3 for the $N=4$ polygon shown in Fig.~\ref{fields} of the main text) by a spin-independent hopping operator $\hat{\mathcal{T}}_\eta$ which is proportional to the hopping energy $t_{\eta,\mathrm{h}}$. The effect of the leads is encoded in the retarded self-energy operators $\hat{\Sigma}_{{\eta}}^{r}(\varepsilon)=\hat{\mathcal{T}}_\eta^\dagger \hat{g}^r_\eta (\varepsilon) \hat{\mathcal{T}}_\eta $, which allow us to obtain the 
retarded Green function, $\hat{\mathcal{G}}^{r}(\varepsilon)$, from the equation $(\varepsilon+\mathrm{i}0^+-\hat{\mathcal{H}}_{p}-\hat{\Sigma}_{\mathrm{d}}^{r}(\varepsilon)-\hat{\Sigma}_{\mathrm{s}}^{r}(\varepsilon))\cdot \hat{\mathcal{G}}^{r}(\varepsilon)=\mathds{I}$. 
In what follows, for simplicity, we do not explicitly write the dependence with $\varepsilon$ of these operators. The zero-temperature linear conductance is computed using the Landauer-Büttiker formula, 
\begin{equation}
G = \frac{e^{2}}{h} \mathrm{Tr}\left[\hat{\Gamma }^\mathrm{s}\hat{\mathcal{G}}^{r} \hat{\Gamma }^\mathrm{d}\hat{\mathcal{G}}^{a} \right]\,,
\label{conductanceTB}
\end{equation}
where the advanced operators, having superscript $a$, are the adjoint of the retarded operators $\hat{O}^{a}=(\hat{O}^{r})^\dagger$, the lead rates operators have matrix elements $\hat{\Gamma }_{j\sigma ,j^{\prime}\sigma ^{\prime }}^{{\eta}
}\! = \!{\mathrm{i}}(\hat{\Sigma}_{{\eta}}^{r}\! - \!\hat{\Sigma}_{{\eta}%
}^{a})_{j\sigma ,j^{\prime}\sigma ^{\prime }}$, and all the operators inside the trace are evaluated at the Fermi energy, i.e., taking $\varepsilon=E_{\text{F}}$. Importantly, the $G$ obtained from \eqref{conductanceTB} is exact and thus it contains all the interfering paths from the source to the drain.   
 
The simulations presented in Fig.~\ref{Tight} of the main text are obtained as follows. We set $N=4$ and take $E_\mathrm{F}/t_\mathrm{h}\approx 0.2 $. The latter implies that  $\lambda_\mathrm{F}/a_0\sim 13.75$, i.e., the Fermi wavelength is well resolved by the discretization. Besides, since we choose $N_s=512$, the semiclassical regime is ensured as the perimeter, $P=2048 a_0$, is much larger than the Fermi wavelength, $P/\lambda_F\approx 149$. To simulate contact leads with a broad energy bandwidth we set $t_\eta=2.5 t_\mathrm{h}$ and choose $\epsilon_\eta$ to set the working energy at the center of the band. To minimize scattering the intermediate hopping energies are set to the average of the hopping in the lead and in the polygon, i.e., $t_{\eta,\mathrm{h}}=(t_\eta+t_\mathrm{h})/2$. As mentioned in the main text, we perform a Fermi energy average of the conductance in order to discard resonance-induced variations due to finite-size effects. Such average energy window includes the Fermi wavevector range $k_F \in [2\pi(n_0-2)/P,2\pi(n_0+2)/P]$, with $n_0=149$, thus ensuring averaging over several orbital resonances. The ballistic case presented in Fig.~2(a) follows directly by considering all four segments being of equal length, i.e., $L_0=P/4$. Instead, for the case of Fig. 2(b), we also average the conductance over disorder realizations, thus simulating the self-averaging of the disorder that arises when measuring the conductance over the full array of hundreds of squares. Each disorder realization is generated by randomly modifying the segment lengths of the 4-sided polygon, i.e., $L=L_0(1+\delta l)$, where $\delta l$ is taken from a $[-0.1,0.1]$ uniformly-distributed probability density.
\subsection*{Appendix C: Spin Dynamics In Square Loops: Analytical Derivation}
\label{AppC}
\renewcommand{\theequation}{C.\arabic{equation}}
\setcounter{equation}{0}
The spin dynamics for counter-clockwise (CCW) and clockwise (CW) propagating carriers in a full round trip around a square circuit of perimeter $P = 4L$ is determined by the unitary operators $U_{\pm}$ introduced in the main text. The conductance $G_2$ presented in Fig.~\ref{Conductance}(b) of the main text follows from $G_2 = e^2/h (2 - \mathrm{Tr}[\Gamma_0\Gamma_0^\dagger])$, with $\Gamma_0$ the overlapped evolution along time-reversed paths: 
\begin{equation}
\label{totaloperator2}
\begin{gathered}
U_{+}=R_{1 4} R_{43} R_{32} R_{21}\,,
\\
U_{-}=R_{12} R_{23} R_{34} R_{41}\,,
\\
\Gamma_0=\frac{U_++U_-}{2}\,.
\end{gathered}
\end{equation}
Here we focus on the main text case, i.e., $\theta=0$ sample orientation with the Dresselhaus and Zeeman fields' directions chosen as shown in Fig.~\ref{fields} of the main text. As $p_y\propto\sin\gamma$, i.e., each segment angle controls the amplitude and sign of the D110 term, it can be seen that the same spin transformations apply for segments with $\gamma=\pi/4$ or $\gamma=3\pi/4$ ($\gamma=-\pi/4$ or $\gamma=-3\pi/4$) because both have positive (negative) $\sin\gamma$ being identical to $+1/\sqrt{2}$ ($-1/\sqrt{2}$). Taking into account the latter consideration in Eq.~(3), the spin evolution operators for different segments can be grouped as follows:
\begin{equation}
\begin{gathered}
\label{spinoperator2}
R_{+}=R_{41}=R_{12}=R_{32}=R_{43}=\exp \left[-\mathrm{i} \boldsymbol{v}_+ \cdot \boldsymbol{\sigma} L\right],
\\
R_{-}=R_{34}=R_{23}=R_{21}=R_{14}=\exp \left[-\mathrm{i} \boldsymbol{v}_- \cdot \boldsymbol{\sigma} L\right],
\end{gathered}
\end{equation}
where the vectors $\boldsymbol{v}_\pm$ represent the two possible effective fields, which can be written as:
\begin{equation}
\label{effective}
\begin{gathered}
\boldsymbol{v}_\pm=\tilde{k}\hat{\mathbf{v}}_\pm~,~~\tilde{k}\equiv\sqrt{k_{\mathrm{Z}}^2+\frac{1}{2}k_{\mathrm{D}}^2}~~,
\\
~~\hat{\mathbf{v}}_\pm\equiv\frac{1}{\tilde{k}}\left( k_{\mathrm{Z}}\cos\alpha, k_{\mathrm{Z}}\sin \alpha, \pm \frac{1}{\sqrt{2}}k_{\mathrm{D}}\right).
\end{gathered}
\end{equation}
By expanding the matrix exponentials $R_\pm=\exp\left[-\mathrm{i}\boldsymbol{v}_\pm\cdot\boldsymbol{\sigma} L\right]$ one obtains:
\begin{equation}
\label{spinoperator3}
R_\pm=a \sigma_0 +\left[x\sigma_x+y\sigma_y\pm z\sigma_z\right],
\end{equation}
where
 \begin{equation}
\label{definitions}
\begin{gathered}
a=\cos{\tilde{k}L},\\
x=-\mathrm{i}\frac{k_{\mathrm{Z}}\cos\alpha}{\tilde{k}} \sin{\tilde{k}L},\\
y=-\mathrm{i} \frac{k_{\mathrm{Z}}\sin\alpha}{\tilde{k}}  \sin{\tilde{k}L},\\
z=-\frac{\mathrm{i}}{\sqrt{2}} \frac{k_{\mathrm{D}}}{\tilde{k}} \sin{\tilde{k}L},
\end{gathered}
\end{equation} with $a\in \mathbb{R}$ and  $ x, y, z \in \mathbb{C}$. The two possible pairwise products of the $R_\pm$ operators can be written as follows:
\begin{equation}
\begin{gathered}
R_+R_-=A\sigma_0+(B-iC)\sigma_x+(D+iE)\sigma_y,
\\
R_-R_+=A\sigma_0+(B+iC)\sigma_x+(D-iE)\sigma_y,
\\
U_{+}=R_+R_-R_-R_+= (A^2+B^2+C^2+D^2+E^2)\sigma_0+\\2A(B\sigma_x+D\sigma_y)+2(CD+EB)\sigma_z,
\\
U_{-}=R_-R_+R_+R_-= (A^2+B^2+C^2+D^2+E^2)\sigma_0+\\2A(B\sigma_x+D\sigma_y)-2(CD+EB)\sigma_z,
\end{gathered}
\label{EqRpRm}
\end{equation}
where we have introduced the definitions: $A=a^2+x^2+y^2-z^2$, $B=2ax$, $C=2yz$, $D=2ay$ and $E=2xz$. Proceeding to calculate $\Gamma_0$ leads to  
\begin{equation}
\label{totaloperator2}
\begin{gathered}
\Gamma_0=\frac{1}{2}(U_{+}+U_{-})=\\(A^2+B^2+C^2+D^2+E^2)\sigma_0+
\\2A(B\sigma_x+D\sigma_y).
\end{gathered}
\end{equation}
\\
The conductance $G_2$ then becomes:
\begin{equation}
\begin{gathered}
\label{conductance}
G_2 = \frac{e^2}{h} (2- \mathrm{Tr}[\Gamma_0\Gamma_0^\dagger])=
\frac{2e^2}{h}-\\\frac{2e^2}{h} (A^2+B^2+C^2+D^2+E^2)^2+
\\ \frac{2e^2}{h} 4A^2 (B^2+D^2),
\end{gathered}
\end{equation}
which in terms of $k_{\mathrm{Z}}$ and $k_{\mathrm{D}}$ reads:

\begin{equation}
\label{conductance2}
\begin{gathered}
\frac{G_2}{\frac{2e^2}{h}}   = 1-\frac{\left(k_{\mathrm{D}}^2/2+k_{\mathrm{Z}}^2\cos\tilde{k}P\right)^2}{\tilde{k}^4}-
\\\frac{\left(k_{\mathrm{Z}}k_{\mathrm{D}}^2 \sin\frac{\tilde{k}P}{2}+k_{\mathrm{Z}}^3\sin\tilde{k}P\right)^2}{\tilde{k}^6}
\end{gathered}
\end{equation}

As discussed in the main text, these analytical results show that $G_2$ is independent of the in-plane Zeeman field's orientation $\alpha$.
\subsection*{Appendix D: Geometrical Interpretation Of Results}
\label{AppD}
\renewcommand{\theequation}{D.\arabic{equation}}
\setcounter{equation}{0}
\begin{figure}[!ht]
\centering
    \includegraphics[width=0.45\textwidth]{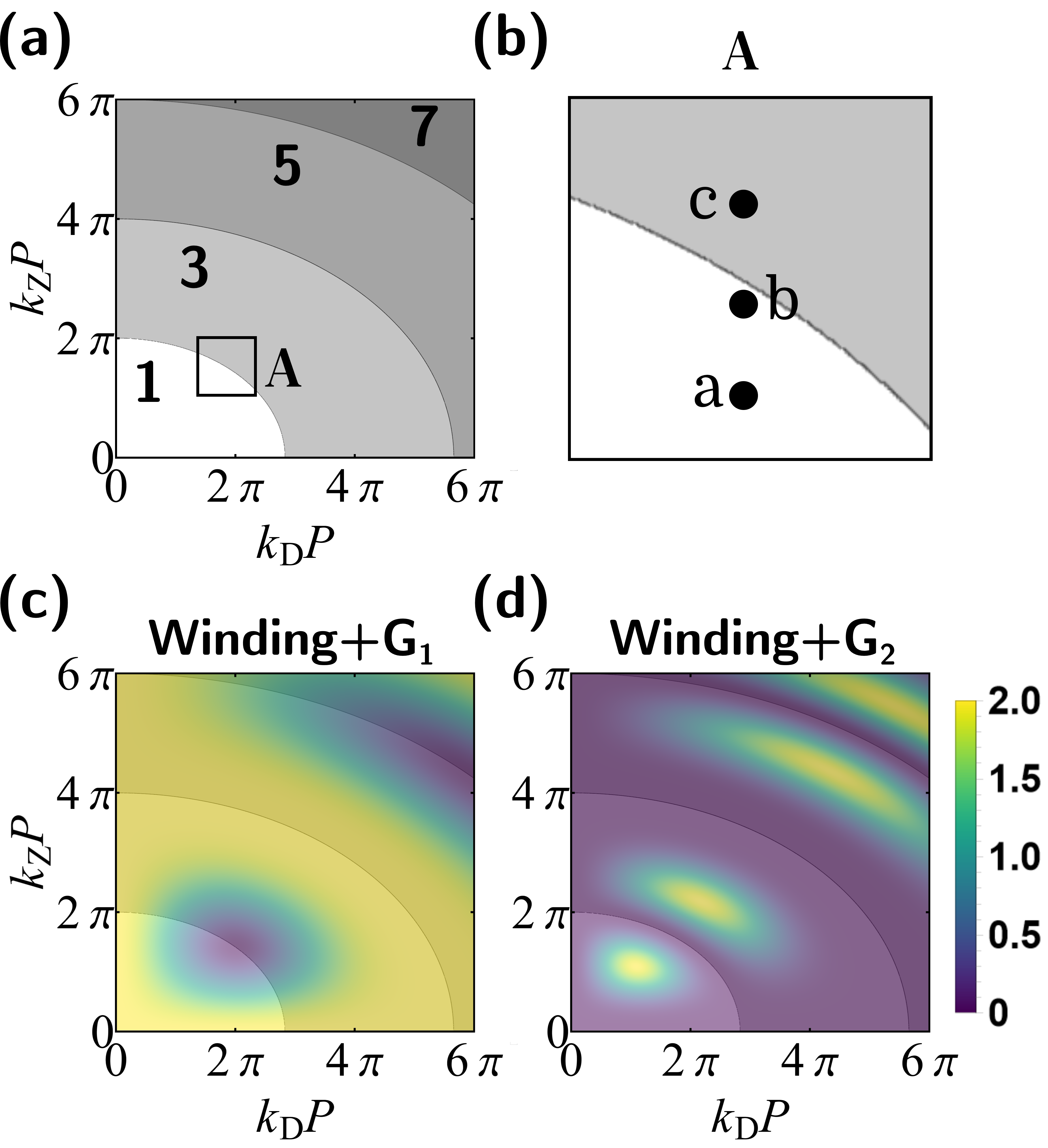}
    \caption{\label{FFWinding}(a) Spin textures' winding number, $\omega$, along the direction of the applied Zeeman field. (b) Zoom on zone A, a boundary between two different topological numbers: see the corresponding spin textures in Fig.~\ref{Bloch}. (c) Winding number overlapped with the conductance $G_1$. (d) Winding number overlapped with the conductance $G_2$.}
\end{figure}
Here, we investigate the relation between the conductance of the square circuit discussed in the main text ($\theta=0$) and the topology of the spin textures determined by the propagating spin modes in a round trip, $|\Psi(\ell)\rangle$, with $\ell\in[0,P]$ a linear parametrization of the circuit's perimeter. 
Such propagating spin modes are found after diagonalizing the CCW evolution operator, $U_+$, and obtaining the eigen-solutions evaluated at vertex $1$, which are then propagated along the sequence of segments using Eq.~\eqref{solutionschr} from the main text. The spin texture of the solution is readily obtained from $\hat{\mathbf{s}}(\ell)=\langle \Psi(\ell)|\boldsymbol{\sigma}|\Psi(\ell)\rangle$, which describes a periodical trajectory on the Bloch-Sphere. The topological characterization of the solutions is made in terms of the winding number of their spin texture around the direction of the Zeeman field \footnote{One can take any of the two CCW or CW solutions to perform this calculation because they all share the same winding number.}. Such winding number is computed as \cite{Wang2019},
\begin{equation}
\label{winding}
\omega=\frac{1}{2 \pi} \int_{0}^{P} d \ell\left(\widehat{\boldsymbol{n}} \times \frac{d \hat{\boldsymbol{n}}}{d \ell}\right) \cdot \hat{\boldsymbol{b}}_\alpha
\end{equation}
where $\hat{\boldsymbol{n}}(\ell)$ is the normalized projection of $\hat{\mathbf{s}}(\ell)$ on the plane orthogonal to the Zeeman field direction $\hat{\boldsymbol{b}}_\alpha$.

Figure ~\ref{FFWinding}(a) shows that the topology of the spin textures as a function of $k_{\mathrm{Z}}P$ and $k_{\mathrm{D}}P$ defines zones of increasing odd winding number. Figure~\ref{FFWinding}(c-d) shows that the boundaries between different topological numbers are correlated with the interference patterns of the conductances. For the conductance $G_2$, as shown in Fig.~\ref{FFWinding}(d), each lobe of maximal conductance is enclosed by a zone of constant winding number. We find that the boundaries between regions with different winding numbers satisfy the condition $U_\pm=\sigma_0$, leading to  $G_2=0$ due to the constructive interference of time-reversed paths. More explicitly, $G_2 = e^2/h (2 - \mathrm{Tr}[\Gamma_0\Gamma_0^\dagger])=0$ since $\Gamma_0=(U_+ +U_-)/2=\sigma_0$. This condition is satisfied whenever $(R_+ R_-)^{-1} = R_- R_+$ which, from Eqs.\eqref{EqRpRm}, requires that $D=B=0$. The latter imposes $\tilde{k}P=2 n\pi$ with $n$ integer, lying exactly at the winding-number boundaries. 

\begin{figure}[H]
\centering
    \includegraphics[width=0.45\textwidth]{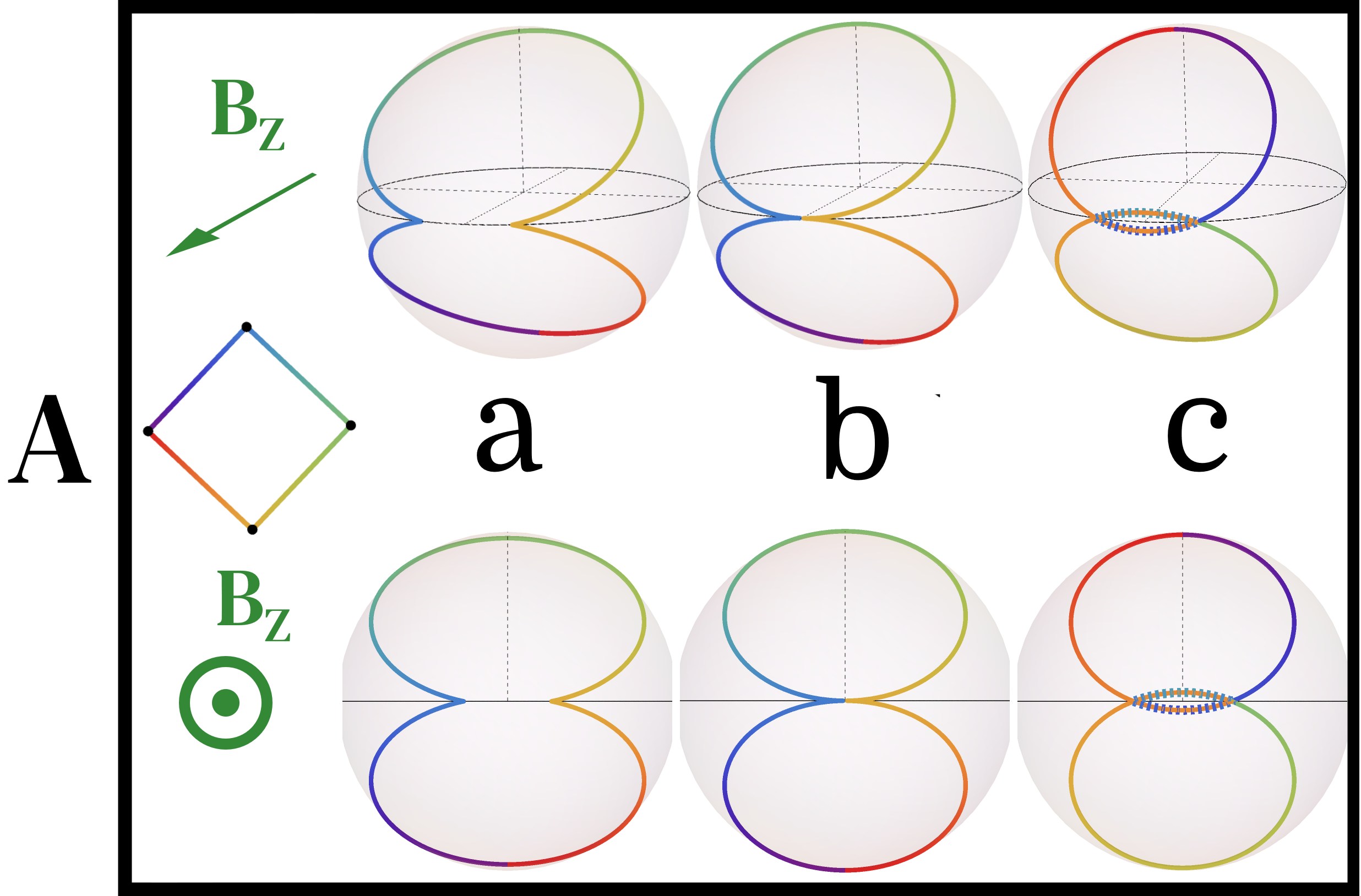}
    \caption{\label{Bloch} (Top) Spin textures of propagating modes in the Bloch sphere for different SO strengths at the zone A shown in Fig.~\ref{FFWinding}(b). (Bottom) Projection of the spin textures on the normal plane to the Zeeman field direction. The color indicates the circulation of the local spin states as the carrier propagates through the perimeter, from red to violet. The insets show the direction of the applied Zeeman field with respect to the Bloch sphere's top and bottom perspectives.}
\end{figure}

In Fig.~\ref{Bloch} we present a series of spin textures undergoing a topological transition along the zone shown in Fig.~\ref{FFWinding}(b). The spin textures in this Dresselhaus 110 square circuit involve two overlapping cones-like trajectories which are seen as eights-like trajectories around the direction of the applied magnetic field. A similar behavior is known to appear in two-level systems subject to a Rabi drive \cite{Zur1983, Stenholm1972, Stenholm1972b}, where the interplay of a constant magnetic field (here the Zeeman field contained in the $xy$-plane) coexisting with a normal oscillating sinusoidal component (here Dresselhaus [110] along the $z$-axis which in this case takes a sequence of discrete values instead of varying continuously) gives rise to spin textures with an analog structure. 

\end{document}